\documentclass[11pt]{article}
\usepackage{graphicx}
\usepackage{amssymb,amsmath}
\usepackage{bm}
\def\lesssim{\;\raise0.3ex\hbox{$<$\kern-0.75em\raise-1.1ex\hbox{$\sim$}}\;}
\def\gtrsim{\;\raise0.3ex\hbox{$>$\kern-0.75em\raise-1.1ex\hbox{$\sim$}}\;}
\def\mdens{{\rm g~cm^{-3}}}

\def\msun{{\rm M}_\odot}
\def\mdot{\dot{M}}

\begin{document}
\baselineskip 21pt
\title{Redshift factor and diffusive equilibrium of unbound neutrons in the single nucleus model of accreting neutron star crust}
\author{P. Haensel and J.L. Zdunik\\
   \\
Nicolaus Copernicus Astronomical Center, Polish Academy of Sciences,\\ 
Bartycka 18, PL-00-716 Warszawa, Poland}
\maketitle
{\it Abstract.}{ Using a Wigner-Seitz approximation with spherical cells, we 
re-analyze  a widely used  single nucleus  model of accreting neutron star crust. 
 We calculate beta disequilibrium within the crust, which is 
 sizable, and implies that neutron and baryon chemical potentials, $\mu_n$ and 
$\mu_{\rm b}$,  are not equal. 
We include also  non-equilibrium reactions, driven by matter compression, and proceeding  in the reaction layers. The  constancy of $e^{\Phi}\mu_n$, 
where the spacetime metric component $g_{00}=e^{2\Phi}$, in the shells between the reaction layers is not applicable,  because single electron captures are blocked, so that 
the neutron fraction is fixed, and therefore  neutrons  are not an independent component of the crust matter. The absence of neutron diffusion  in  the shells  between the reaction layers, stems from the constancy of the neutron fraction (concentration) in these shells. In the reaction layers, the outward force  resulting  from  neutron fraction gradient is balanced by the inward gravitational force acting on unbound neutrons. Neglecting the thickness of the reaction layers  compared to the shell thickness, we obtain  condition  $e^{\Phi(r)}f_Q(r)g(r)=$constant, where $g$ is Gibbs energy per nucleon, undergoing  discontinuous drops on the reaction surfaces,  and
$f_Q(r)g(r)={\widetilde{g}}(r)$ is a continuous function, due to the factor 
$f_Q(r)$ canceling   the discontinuities (drops) in  $g(r)$. The function $f_Q(r)$ 
is calculated using the Tolman-Oppenheimer-Volkov equations from $f_Q(P)$ and 
$g(P)$ obtained from the equation of state (EOS) with discontinuites. The constancy of 
of $e^{\Phi(r)}{\widetilde{g}}(r)$ is an extension of the  standard 
relation  $e^{\Phi(r)}{{\mu}}_{\rm b}(r)=$constant,  valid in hydrostatic equilibrium for catalyzed 
crust.}
\section{Introduction}
\label{sect:Introduction}
A model of the matter with only one nuclear species present at a given pressure 
yields  a simplest approximation  of the neutron star crust, called frequently a single nucleus model (SNM). The crust is treated 
there as a one component plasma.
{ The popularity of the SNM stems from its simplicity. }
SNM was used  in the calculation of the 
accreted crust structure \cite{Sato1979,HZ1990b} and crustal heating in accreting neutron stars, generated by the non-equilibrium nuclear processes induced by the compression of the crust matter \cite{HZ1990a,HZ2003,HZ2008}. { In spite of its simplicity and some 
seemingly unrealistic features}, 
including sharply localized heat sources, SNM yields cumulated deep crustal heating consistent with advanced numerical simulations
involving multi-component plasma and large nuclear reaction networks 
\cite{Gupta2007,Gupta2008,LauBeard2018}. 
 Recently, the  validity of the SNM for calculating the crust structure and 
the cumulated deep crust heating in accreting neutron stars  has been questioned \cite{Fortin2019}.  This was argued to result in neutron diffusion, and a strong decrease of the deep crustal heating. 

In the present paper we address the above mentioned problems. 
 { Additionally, we show using the Tolman-Oppenheimer-Volkov (TOV) equations, that for Gibbs energy per nucleon $g(r)$ with 
discontinuous drops associated with heat release, 
the standard constancy relation $e^{\Phi(r)}{{\mu}}_{\rm b}(r)={\rm constant}$ 
is replaced by 
$e^{\Phi(r)}{\widetilde{g}}(r)=constant$, with ${\widetilde{g}}(r)=
f_Q(r)g(r)$. Here, $f_Q(r)$ is a step-like function making $
{\widetilde{g}}(r)$ continuous, calculated using the TOV equation 
that yield pressure profile $P(r)$ and $\Phi(r)$ for an assumed EOS.} The constancy of $e^{\Phi(r)}{\widetilde{g}}(r)$ holds throughout  both the outer (no free neutrons, but still $g(r)$ discontinuities present!) and the inner crust. 

In Sect.2 we review essential features of the SNM of the crust, 
and applications of this model to the simulations of an 
accreting neutron star crust.  We point out the differences  between the accreted 
and catalyzed crusts, and illustrate the applications of the SNM to both 
cases, using a  nuclear model for the nucleon component of the matter. 
In the last subsection of Sect.2 we demonstrate  { diffusive equilibrium of unbound
neutrons in } the accreted crust.  

In Sect.3 we consider the hydrostatic equilibrium of an accreted crust and
derive an extension of $e^{\Phi(r)}{{\mu}}_{\rm b}(r)={\rm constant}$ theorem 
to the case of of $g(r)$ with discontinuities (sharp drops) characteristic
of SNM of accreting neutron star crust. Section 4 presents discussion of our results and
conclusion. 
\section{Single-nucleus model of the inner accreted crust}
\label{eq:beta-single.nuc}
In the present paper we obtain  strong and electromagnetic interaction 
equilibrium { of the crust} by putting $Z$ protons and electrons,  and 
$N_{\rm cell}=A_{\rm cell}-Z$ neutrons into  a spherical Wigner-Seitz (W-S)  cell under pressure $P$ and  calculating the ground state of the system using an approximate solution 
 of the nuclear  many-body theory. Temperature effects are neglected and $T=0$ 
 approximation is used. The ground state has a proton cluster at the cell center,
{ neutrons bound to the proton cluster, and  beyond neutron drip pressure, $P_{\rm ND}$,}  also a  fraction of $N_{\rm cell}$ unbound and filling the W-S cell. The possibility of the non-spherical pasta phases at the bottom of the crust will not be  considered. By construction, $A_{\rm cell}$ is an integer number. We  also calculate  the neutron chemical potential 
$\mu_n$, as well as $\mu_p$ and $\mu_e$ (all $\mu$-s include rest energies of 
particles). { The Gibbs  free energy  per nucleon} (equal to  baryon chemical 
potential $\mu_{\rm b}$)   is 
\begin{equation}
g=[Z(\mu_p+\mu_e)+N_{\rm cell}\mu_n]/A_{\rm cell}=
x_p(\mu_p+\mu_e)+(1-x_p)\mu_n~,
\label{eq:g1.single}
\end{equation}
where the proton fraction $x_p=Z/A_{\rm cell}$.  
Expression for $g$, Eq.(\ref{eq:g1.single}), can be rewritten as
\begin{equation}
g=x_p(\mu_p+\mu_e-\mu_n)+\mu_n=x_p\Delta\mu+\mu_n~.
\label{eq:g2.single}
\end{equation}
A strict beta equilibrium  between $n$, $p$, and $e$ corresponds to $\Delta\mu=0$  and 
$\mu_{\rm b}=\mu_n$. 
\subsection{Accreted crust}
\label{sec:AC}
For the sake of simplicity, we will limit to the case of a 
stationary fully accreted crust. The structure of such a 
crust, calculated using the SNM, was 
derived  in numerous papers \cite{Sato1979,HZ1990b,HZ1990a,HZ2003,HZ2008}. 
It is obtained  by simulating a 
compressional evolution  of a  W-S cell $(A_{\rm cell},Z;P)$ with 
$P$ increasing due to the weight of accreted matter, and taking 
into account possible electron captures, neutron emissions and 
absorptions, and at density $\rho\gtrsim 10^{12}~\mdens$, also pycnonuclear 
fusion of neighbouring nucleon clusters, driven by the  quantum zero point
oscillations  of the clusters and penetration of the Coulomb barriers 
between them. 

In the process of formation of a fully accreted crust, nuclear ashes  at $\sim 
10^8~\mdens$, resulting from an explosive  thermonuclear burning of a freshly accreted 
plasma which generates the X-ray bursts in a LMXB, are compressed up to $10^{14}~\mdens$ after reaching the bottom of the crust. 
This compressional evolution can be followed within the SNM  by simulating compression of 
a single initial W-S cell $(A_{\rm cell}^{\rm (in)},Z^{\rm (in)})$ under 
increasing pressure $P$,  inducing electron captures decreasing $Z$ at 
constant $A_{\rm cell}$, then neutron drip combined with electron captures for densities exceeding $5\times 10^{11}~\mdens$, still at constant $A_{\rm cell}=A_{\rm cell}^{\rm (in)}$, and finally including also pycnonuclear fusions accompanied by the 
electron captures and neutron emissions and absorptions, resulting in a doubling of 
$A_{\rm cell}$. As a final result, one gets an evolutionary track given as  
$(A_{\rm cell},Z)_{\rm AC}$ as a function of $P$. At $T=0$ reactions start  
at specific threshold values  $P=P_j$. The evolutionary trajectory  in the 
$A_{\rm cell},Z$ plane follows the local minimum of $g(A_{\rm cell},Z;P)$. 
The minimization is done at fixed $A_{\rm cell}$, with an option $A_{\rm cell}\longrightarrow 2A_{\rm cell}$ open if the timescale of the pycnonuclear fusion (usually
taking place after electron captures) becomes shorter than a local compression 
timescale. 

Let us remind, that in the case of the crust made of cold catalysed matter, its composition calculated using the SNM corresponds to an absolute  (global) 
minimum at a given $P$, i.e., to the ground state (GS), and complete 
thermodynamic equilibrium  of the crust $(A_{\rm cell},Z;P)_{\rm GS}$. 
The track  $(A_{\rm cell},Z)_{\rm GS}$ is very different from the 
$(A_{\rm cell},Z)_{\rm AC}$ one, with $Z_{\rm AC}(P)$ values being significantly 
lower  {(by a factor  $\sim 2-3$ )} than $Z_{\rm GS}(P)$. 

Both GS and AC crusts have an onion-like structure consisting of shells 
$(A^{(j)},Z^{(j)})$, $j=1,\ldots,j_{\rm max}$. The transition pressure $P_j$ between $j$ and $j+1$ shell
is associated with a density discontinuity (drop) much smaller in the 
GS crust than in the AC crust.  As $P_j^{\rm AC}$ corresponds to the threshold
for an equilibrium electron capture followed by a second non-equilibrium  electron capture, and possibly  emission of neutrons, there is not only a density 
drop at $P_j^{\rm AC}$, but also a $\Delta g(P_j^{\rm AC})$ drop associated with a heat release per one
accreted nucleon.

Another  important feature of an AC crust, differing it from the GS crust,  is 
a (significant) metastability of the local equilibrium state, between the consecutive reaction surfaces $P_i$, where $P_i$ is a threshold pressure for a single electron
capture initiating transition to a different (lower) local minimum of $g$. 
For $P_{i-1}<P<P_{i}$ the W-S cell is in a metastable state, because  single electron capture is blocked by the energy barrier, while the double electron capture -  
obviously leading to a lower $g$ because of the nucleon paring -  is  assumed to be too  
slow to proceed.

 The integer parameters $Z$ and $A_{\rm cell}$ (and therefore  neutron  and proton fractions within nucleons) stay constant  within shells $P_{i-1}<P<P_i$. 
At each $P=P_i$, the value of $\mu_{\rm b}$ undergoes a drop by $Q_i$, while 
$P$ changes from $P^-_i=P_i-0$ to $P^+_i=P_i+0$. So $P_i$ determines a surface
of discontinuity of $\mu_{\rm b}$. This spherical surface is an idealization 
of the heating (reaction) layer in an accreting crust.   

\begin{figure}[h]
\resizebox{\hsize}{!}{\includegraphics{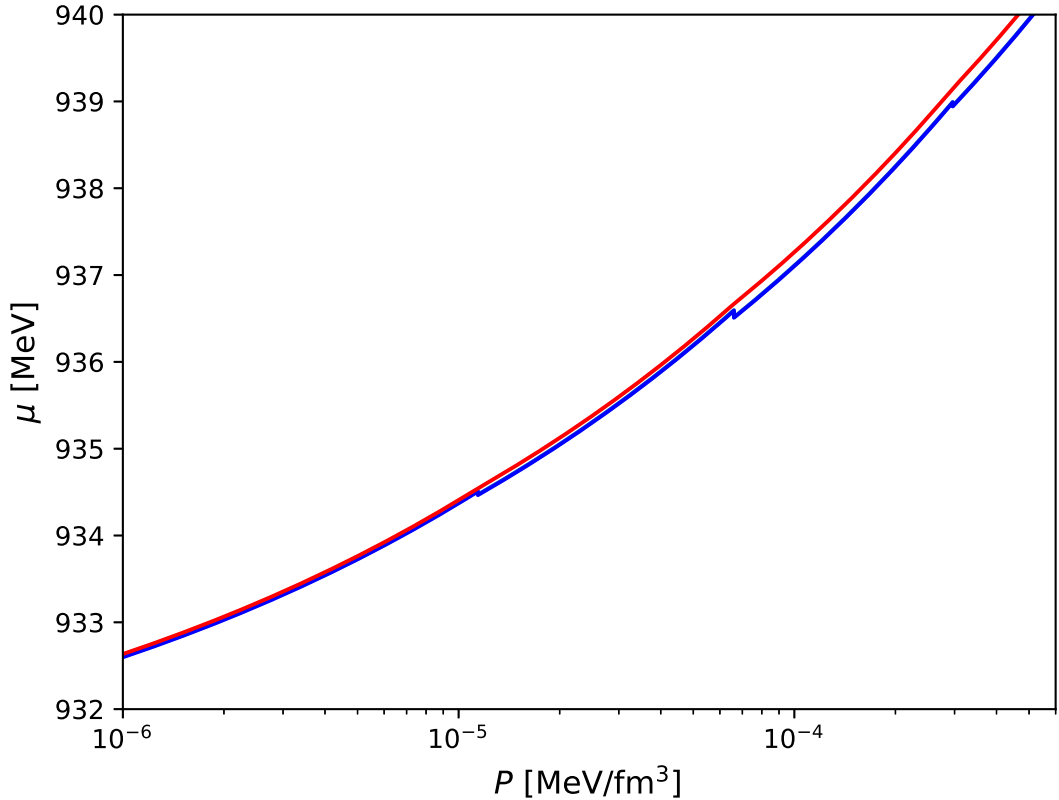}}
\caption{Baryon chemical potential  $\mu_{\rm b}$ versus pressure (blue line) in the outer
crust. The drops in $\mu_{\rm b}$ correspond to thin heating layers. In red - 
$\widetilde{g}(P)$ defined in Sect.\;\ref{sect:n-AC}.  Calculations performed for the MB model of the nucleon sector. }
\label{fig:mub.P.outer}
\end{figure}

\begin{figure}[h]
\resizebox{\hsize}{!}{\includegraphics{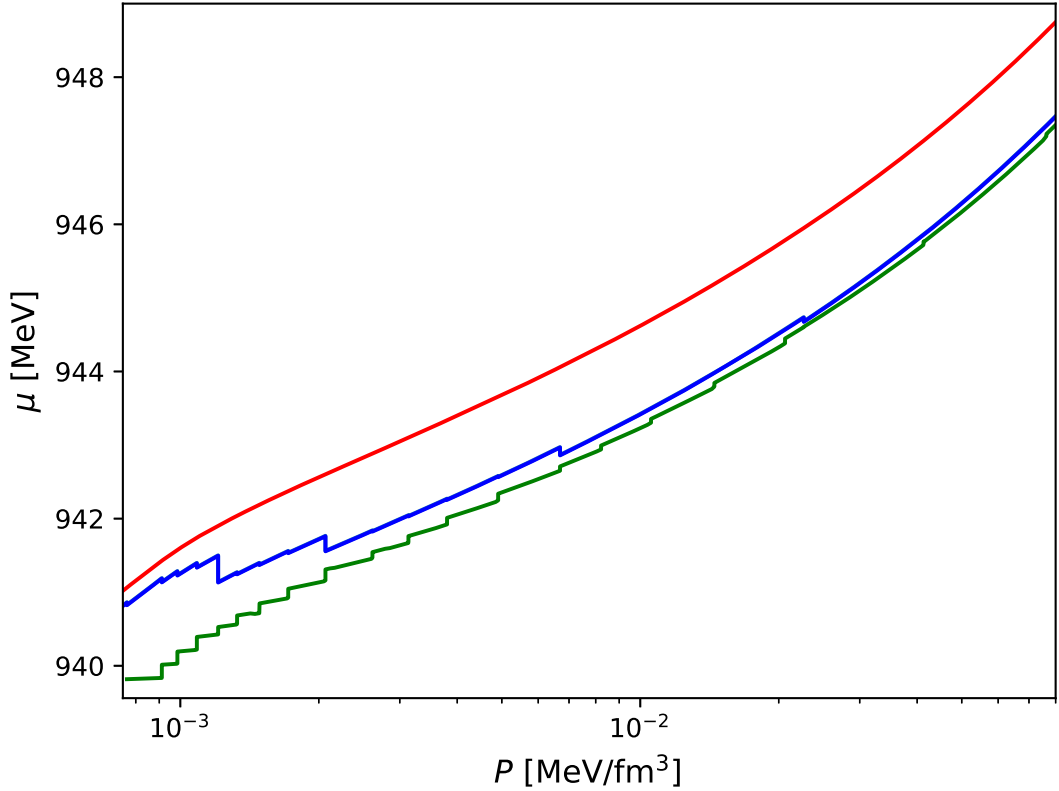}}
\caption{Baryon chemical potential  $\mu_{\rm b}$ versus pressure (blue line) in the 
inner crust. The drops 
in $\mu_{\rm b}$ correspond to thin heating layers. In red - 
$\widetilde{g}(P)$ defined in Sect.\;\ref{sect:n-AC}. Green curve - $\mu_{\rm no}$ - neutron chemical potential of the neutron gas. Calculations performed for the MB model of the nucleon sector.
 }
\label{fig:mub.P.inner}
\end{figure}
Specific features  of an  EOS.AC of a fully accreted crust are illustrated in 
Figs.\;\ref{fig:mub.P.outer},\ref{fig:mub.P.inner}. The calculations were done 
for Mackie-Baym (MB) model of the nucleon sector of the crust plasma 
\cite{MackieBaym}. 
\subsection{Diffusive equilibrium  of  unbound neutrons  in the inner accreted crust}
\label{sec:n-diff}
Protons cannot diffuse because they are bound (confined) 
in the localized clusters, which in turn are localized  at the lattice 
sites of the Coulomb crystal. The electrons are coupled to protons 
by the electromagnetic forces. 
Neutrons do not diffuse within shells between reaction 
surfaces, because neutron fraction per nucleon within a 
$j-$th shell,  $x_{n}^{(j)}=1-x_p^{(j)}$, is constant, 
${\rm d}x_n^{(j)}/{\rm d}r = 0~$  \cite{LanLif-FM} (we neglect corrections of the 
order of $(R-r)/R$, where $R$ is the NS radius). 
In reality, the  changes of 
 matter composition due to  reactions triggered by the electron capture 
take place in a layer separating the two shells. In this 
reaction layer we are dealing with a  two-component plasma of proton 
clusters, $Z^{(j)}$ and $Z^{(j+1)}$, neutrons, and electrons.  
The neutron fraction per nucleon 
grows with depth $z=R-r$ from $x_n^{(j-1)}$ to  $x_n^{(j)}$, { 
and  ${\rm d}x_n/{\rm d}z> 0~$, but resulting generalized force  
${\rm d}\mu_n/{\rm d}z$ acting outwards is balanced by the gravitational 
pull  $g_z m_n$. In the Newtonian approximation
\begin{equation}
{{{\rm d}\mu_n}\over{{\rm d}z }}= m_n g_z
\label{eq:diff-n.1}
\end{equation}
which can be integrated over the reaction layer $[z_{j},z_{j+1}]$,} 
\begin{equation}
\mu_n(z)= m_n g_r\cdot (z-z_{j})+\mu_n^{(j)}~,
\label{eq:diff-n.2}
\end{equation}
where  $\mu_n^{i}=\mu_n(P_i)$, $i=j,j+1$
This is diffusive equilibrium  condition within the $j$th reaction 
layer between $j$th and $j+1$th shells: neutron diffusion outwards is blocked by the
gravitational pull acting on  a neutron  inwards.   

Denoting by $d_j$ the thickness of the $j$th reaction layer, 
we get from Eq.\;(\ref{eq:diff-n.2}) an estimate
\begin{equation}
d_j= {\Delta\mu_n^{j}\over {m_n g_z}}\simeq 10~{\rm m}~,
\label{eq:diff-n.3}
\end{equation}
where  $\Delta\mu_n^{j}=\mu_n^{j+1}-\mu_n^{j}$.
This allows for an estimate of a time for an element of matter to cross the
reaction layer, due to accretion onto the NS surface,
\begin{equation}
\tau_{\rm cross}=4\pi r^2\rho\; {d_j/\mdot}~.
\label{eq:tau.cross}
\end{equation}

 In the reasoning presented above we did not consider 
timescale of neutron diffusion process stemming from the 
scattering of neutrons on nucleon clusters ("nuclei"), $\tau_{\rm diff}$, 
and its interplay with two other timescales, $\tau_{\rm cross}$, 
and single electron capture timescale, $\tau_{\rm cap}$, 
which strongly depends on $\mu_e(z)-W$ where $W$ is the energy 
threshold. Self-consistent treatment  of all three processes -  unbound neutron 
diffusion, electron capture, and the inward matter flow due to accretion  
is beyond the scope of the present paper, and will be presented in our forthcoming 
publication.
  
  In the simplest version of the SNM  the reaction layer is 
replaced by a surface with appropriate boundary conditions on  
both its sides. Actually, SNM of accreted crust can be extended to 
treat a finite thickness of the reaction layer \cite{Bildsten1998a}. 
Using the $\mu_n(r)$ profile  and the reaction layer thickness $\Delta r$, 
one can estimate  heating intensity ${\rm d}Q^{(j)}/{\rm d}r$ profile
within $j$-th reaction layer, and then obtain the cumulated deep crustal 
heating
\begin{equation}
Q(r)=\int_{r_1^{(+)}}^{r}{\rm d}r {\sum_{(j)}{{\rm d}Q^{(j)}\over{{\rm d}r}}}
= \int_{P_1^{(+)}}^{P}{\rm d}P {\sum_{(j)}{{\rm d}Q^{(j)}\over{{\rm d}P}}}.
\label{eq:Q.cum}
\end{equation}
The cumulated heating due to continuous sources is reasonably well 
reproduced by  the simple SNM with { infinitely} thin sources and 
${{{\rm d}Q^{(j)}}/{{\rm d}r}}=Q_j\delta(r-r_j)$ \cite{HZ2008}. 

The commonly used diffusive equilibrium  condition $e^\Phi\mu_n=$constant 
applies only when  neutrons constitute an independent component of the 
crust  \cite{LanLif-SP1}. This is not the case for the shells between the reaction layers, where beta processes are blocked and $x_n=1-x_p$ is fixed. Within  the shells 
there  is only one independent density, $n_{\rm b}(P)$, and the corresponding chemical 
potential  is $\mu_{\rm b}(P)$ (see next section). To determine thermodynamic 
equilibrium in a $j$-th shell with fixed $x_n^{(j)}$ it is sufficient to know
$P$ or $n_{\rm b}$.  In the next section, an extension of $e^\Phi\mu_n=$constant
 to the case of accreted crust, is derived from the 
 TOV equation  relating  equation of state (EOS) of accreted crust  and  $\Phi(r)$.  

Diffusion of unbound neutrons in neutron star crust was studied in \cite{Bisno1976} 
in the context of heating of the envelopes of {\it single}  X-ray sources. An initial 
 large excess of unbound neutrons was a leftover from a hot  state of a newly born neutron star, where a significant fraction of unbound neutrons  existed  at densities below cold neutron drip point (see, e.g., Fig.\;3.1 in \cite{NSB}). A rapid cooling led to formation of a non-equilibrium layer with a sizable neutron   excess. Neutron diffusion inwards was then driven mostly by the { gravitational pull}  and generated heating of the crust during some $10^4$ yr after which neutron  equilibrium was reached. The scenario considered  in \cite{Bisno1976} is different from that associated  with accreting neutron stars. 
\section{Hydrostatic equilibrium, EOS, and metric function $\Phi$}
\label{sect:mu_n-TOV}
The TOV equation for  $\Phi(r)$   is 
\begin{equation}
{{\rm d}\Phi\over {\rm d}r}=-{1\over {\cal E} + P}{{\rm d}P\over {\rm d}r}~,
\label{eq:Phi}
\end{equation}
where ${\cal E}$ is the energy density of the matter (including rest energies
of the matter constituents). 
The dimensionless  pseudoenthalpy $H(P)$ is defined by (e.g., \cite{NSB})
\begin{equation}
H(P)=\int_0^P {{\rm d}P^\prime\over {{\cal E}(P^\prime)+P^\prime}}~.
\label{eq:def-H}
\end{equation}  
As we assume that  remaining  TOV equations have been integrated, we 
can use $P(r)$ profile corresponding to the hydrostatic equilibrium 
of NS.  Notice that $H(P)$ and  $P(r)$ are  smooth (differentiable), even when  
${\cal E}(P)$  and $n_{\rm b}(r)$ are not. Therefore, { we can rewrite 
Eq.(\ref{eq:def-H})}   as 
 \begin{equation}
{{\rm d}H\over {\rm d}r}={1\over {{\cal E}(P)+P}}{{\rm d}P\over {\rm d}r}~, 
\label{eq:der-H}
\end{equation}  
so that  
 \begin{equation}
{{\rm d}\over {\rm d}r} \left[ H(r)+\Phi(r)\right]=0~, 
\label{eq:Bern1}
\end{equation}  
which results in 
 \begin{equation}
H(r)+\Phi(r)={\rm constant}=\Phi(R)~,    
\label{eq:Bern2}
\end{equation}  
where $R$ is the radius of NS. 

In the case of complete  { thermodynamic } 
equilibrium (cold catalyzed matter)  $g=\mu_{\rm b}$ 
and we obtain well known constancy of $e^{\Phi(r)}\mu_{\rm b}(r)$ \cite{Harrison1965}, 
valid within NS built of cold catalyzed matter. 
Simultaneously, $\mu_{\rm b}$ can be replaced by $\mu_n$ (Sect.\;2), resulting in  equilibrium condition  for neutrons ${\rm e}^{\Phi(r)}\mu_{n}(r)=$constant.
However, an accreted  crust is   off beta equilibrium. Moreover, simplest version 
of SNM yields sharp discontinuities in $\mu_{\rm b}$ and $\mu_n$.

Before reconsidering the problems related to the SNM of accreted crust, let us 
remind the derivation of ${e^{\Phi}}\mu_n=constant$ for uniform $npe$ matter in beta
equilibrium. In this case Gibbs free energy per baryon $g(P)=({\cal E}+P)/n_{\rm b}=\mu_{\rm b}=\mu_n$,  
 ${\rm d}P=n_{\rm b}{\rm d}g$, 
 and the integral in $H(P)$ can be taken to yield
 \begin{equation}
H(P)={\rm \ln}\left[{g(P)\over g(0)}\right]~.  
\label{eq:beta-eq}
\end{equation}  
Then Eq.(\ref{eq:Bern2}) could be written as
 \begin{equation}
{\mu_n(r) {\rm e}^{\Phi(r)}={\rm constant}},  
\label{eq:mu_n-beta}
\end{equation}  
valid over the whole core region composed of $npe$ matter in beta 
equilibrium.
\subsection{Accreting  neutron star crust}
\label{sect:n-AC}
Let us consider a  simplest SNM of a fully accreted crust. 
It consists of shells with fixed $(A_{\rm cell},Z)_{\rm AC}$, separated 
from the neighbouring shells by spherical surfaces  $P=P_{i-1},~P_i$. While
$P$ is continuous accross these surfaces, $g, {\cal E }, n_{\rm b}$ suffer
a discontinuity there. Therefore we have to precisely define the boundary conditions
on both sides of the $P_i$ surfaces. Let $P^+_{i-1}=P_{i-1}+0$, $P_i^-=P_i$. Within
$P_{i-1}^+\leq P \leq P^-_i$  function $g(P)$ is continuous.  Define also $g_j=g(P^-_j)$. Then  $H(P)$ can be 
split into a sum of integrals of continuous functions,
\begin{equation}
H(P)=\int_0^{P^-_1}{{\rm d}P^\prime\over {n_{\rm b}(P^\prime)g(P^\prime)}} + 
\int_{P^+_1}^{P^-_2}{{\rm d}P^\prime\over {  n_{\rm b}(P^\prime)g(P^\prime)}} + \ldots
+\int_{P^+_j}^{P}{{{\rm d}P^\prime}\over { n_{\rm b}(P^\prime)g(P^\prime)}}~.
\label{eq:H.sum}
\end{equation}
Relation between ${\rm d}P$ and ${\rm d}g$ for a cell with fixed $A_{\rm cell}$ is
\begin{equation}
{\rm d}g={\rm d}P/n_{\rm b}+\sum_{i=e,n,p}\mu_i{\rm d}x_i~.
\end{equation}
As on the continuous segments of $g(P)$ fractions  $x_i$ are constant, we get 
${\rm d}P=n_{\rm b}{\rm d}g$, and all integrations in 
Eq.\;(\ref{eq:H.sum}) are easily taken, leading to 
\begin{equation}
H(P)={\ln}\left[\frac{g(P)}{g_0}\cdot {g_1\over g_1-Q_1}\cdot \ldots  
\cdot {{g_j}\over{ g_j -Q_j}}\right]~,
\label{eq:H.P.Qj}
\end{equation}
where $Q_j$ is energy release per one accreted nucleon  on  the $j$-th 
reaction surface and  $P_j<P<P_{j+1}<P_{j_{\rm max}}$. {
Equation (\ref{eq:H.P.Qj}) can be rewritten in the form 
\begin{equation}
H(P)=\ln\left[{g(P)\over g_0}\prod_{j=1}^{j_{\rm max}}
\left({g_j\over {g_j-Q_j}}\right)^{\Theta(P-P_j)}\right],
\label{eq:P.H.prod}
\end{equation}
where the Heaviside  function $\Theta(x)=1$ for $x>0$ and $\Theta(x)=0$ otherwise.
Let us denote
\begin{equation}
f_Q(P)=\prod_{j=1}^{j_{\rm max}}
\left({g_j\over {g_j-Q_j}}\right)^{\Theta(P-P_j)}~.
\label{eq:fQ.P}
\end{equation}
The product of two discontinuous functions, $g(P)$ and $f_Q(P)$, 
is a continuous function $\widetilde{g}$,
\begin{equation}
\widetilde{g}(P)=f_Q(P)g(P)~.
\label{eq:tilde.g}
\end{equation}
We can call $\widetilde{g}(P)$ as regularized EOS for accreted crust. It is 
represented by red lines in Figs.\;\ref{fig:mub.P.outer},\ref{fig:mub.P.inner}.
It was already introduced in \cite{HZ2003} as a continuous envelope of a set 
of discountinuous EOS's for accreted crust. It was demonstrated  in
\cite{ZFH2017} that function $f_Q(P)$ describes relative difference in thickness of the accreted and catalyzed crust.
}

Using our  expression for $H(P)$, together with 
pressure profile $P(r)$ within the hydrostatic static configuration
of NS, we can rewrite Eq.(\ref{eq:Bern2}) in the form
\begin{equation}
\Phi(r)+H(r)=\ln g(r)-\ln g_0 + 
\sum_{j=1}^{j_{\rm max}}
\ln\left({g_j\over {g_j-Q_j}}\right)\Theta(r-r_j)+\ln(e^{\Phi(r)})-\ln(e^{\Phi_0})~,
\label{eq:PhiH}
\end{equation}
where  $\Phi_0=\Phi(R)$ and  $g_0=g(R)$. 
This leads to 
\begin{equation}
f_Q(r) g(r)e^{\Phi(r)}={\rm constant}=g(R) e^{\Phi(R)}~,
\label{eq:gQconst}
\end{equation}
where 
\begin{equation}
f_Q(r)=\prod_{j=1}^{j_{\rm max}}\left({g_j\over {g_j-Q_j}}\right)^{\Theta(r-r_j)}~.
\label{eq:fQ}
\end{equation}
The above condition is an extention of  the standard relation for a one-parameter EOS 
$\mu_{\rm b}e^\Phi=$constant with a continuous $\mu_{\rm b}(r)$, derived 
for cold catalyzed matter \cite{Harrison1965}. In the strict beta equilibrium $\mu_{\rm b}=\mu_n$ and with  all $Q_i$ vanishing, we get  $\mu_n e^\Phi=$constant. For the SNM of fully accreted crust,  $g(r)$ undergoes discontinuous drops  at $r_j$, and  
a continuous ${\widetilde{g}}(r)=
f_Q(r)\mu_{\rm b}(r)$, fulfills  
$e^{\Phi(r)}{\widetilde{g}}(r)=$constant.
\begin{figure}[h]
\resizebox{\hsize}{!}{\includegraphics{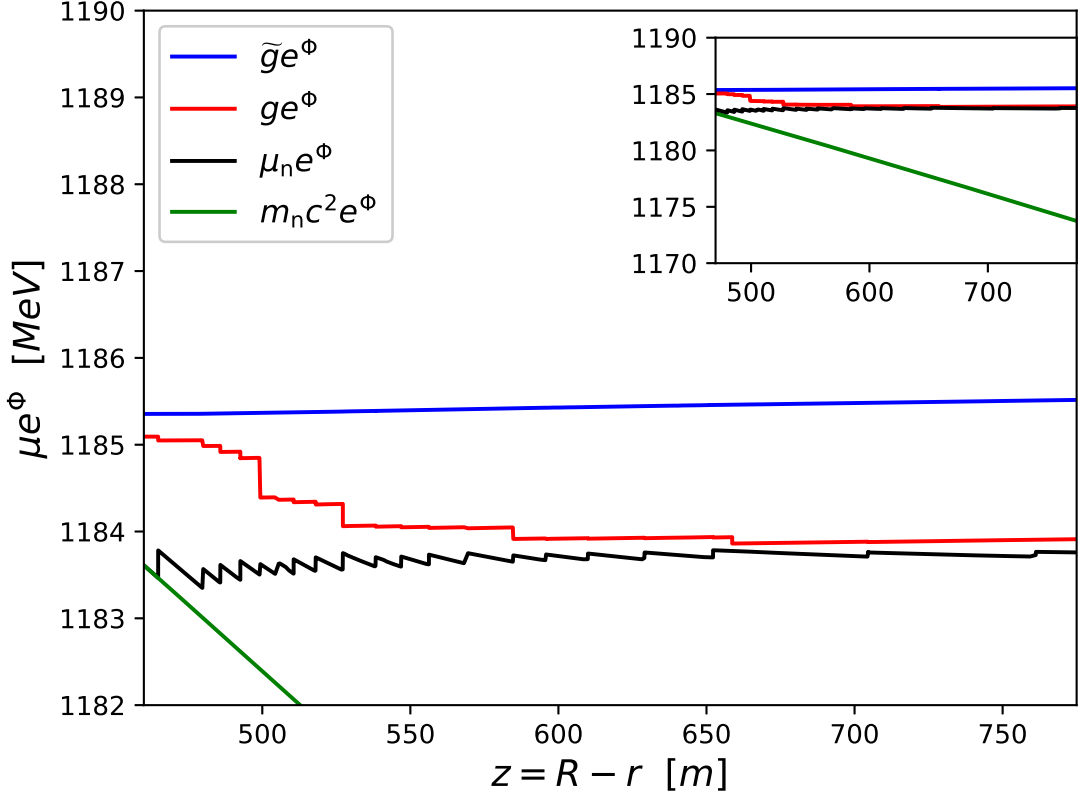}}
\caption{ The quantities  $e^{\Phi(z)}\mu_n(z)$ (black line),
 $e^{\Phi(z)}g(z)$ (red line) and 
$e^{\Phi(z)}{\widetilde{g}}(z)$ (blue line) vs. depth $z=R-r$ within  a 
fully accreted inner crust for NS mass $1.4 \;\msun$.
 Calculations performed for the MB model of the nucleon sector. A barely visible minute deviation  of the $e^{\Phi(z)}{\widetilde{g}}(z)$ line from constancy results 
 from limitations of numerical precision of solving relevant nonlinear equations. 
 The lowest (green) line   $m_n c^2 e^{\Phi(z)}$ shows the overwhelming 
 importance of the z-dependence of the redshift factor. 
 }
 \label{fig:tilde.g.mun}
 \end{figure}

 Deviation of $e^{\Phi(r)}\mu_n(r)$ from constancy in fully accreted inner crust
fluctuates around mean value with maximum amplitude $\sim  0.4\,{\rm MeV}$ reached 
after the neutron drip point,  and then  decreasing with density and depth. This is to be contrasted with constancy of 
$e^{\Phi(r)}\widetilde{g}(r)$, as visualized in Fig.\;\ref{fig:tilde.g.mun}.

The case of the accreted outer crust deserves a comment. There, all $N_{\rm cell}$ neutrons are bound in nuclei, so that $\mu_n<m_n c^2$, where $m_n$ is the neutron mass in 
vacuum. Beta equilibrium is not fulfilled, and $\mu_{\rm b}\neq \mu_n$. 
Let us notice, that a minor
breaking of equality $\mu_{\rm b}=\mu_n$  occurrs also in the SNM of cold catalyzed matter, 
and results there from the discreteness of $N$ and $Z$. This  has been noted 
in the classical paper \cite{BPS1971}. 

Let us also notice that for accreted crusts  $Q_j/g_j\lesssim 10^{-3}$. Therefore 
an approximation of the regularizing  factor $f_Q$ in Eq.(\ref{eq:gQconst}) 
by $1 + \sum_i Q_i/g_i \Theta(P>P^+_j)$ is precise within  $10^{-5}$. 
\section{Discussion and conclusion}
We reconsidered the simplest version of the  single 
nucleus model for fully accreted NS crust, 
with reactions induced by  matter compression,  acting on reaction surfaces determined 
by the electron capture threshold. 
As we have shown, this simplest SNM  does not violate   neither the 
general  relativistic  equations of hydrostatic equilibrium, nor diffusive equilibrium 
of unbound neutrons.  It fulfils a generalized constancy condition 
$e^\Phi \widetilde{g}=$constant where $\widetilde{g}$ is $f_Q\cdot g$ with regularizing 
factor $f_Q$  calculated from the EOS with discontinuities.
We considered  also a more realistic SNM with finite thickness of the reaction layer \cite{Bildsten1998}, and we have found diffusive equilibrium of the unbound neutrons within this layer. 

The simplest SNM of accreted crust looks very crude, and it does not represent 
details of the deep crust heating. The step-like cumulated (integrated) heating 
looks very schematic. However, it is sufficient to model the thermal 
state of NS in LMXB in quiescence. 

Some extentions of the SNM are to be made. This includes modeling of the 
finite thickness of the reaction layers \cite{Bildsten1998a}, 
with possible overlapping of them, and inclusion of the temperature effects, 
that allow for the sub-threshold electron captures \cite{Bildsten1998,Bildsten1998a}. 
We do not think that these extensions of the SNM will affect
the integrated heat and therefore the predicted thermal state of NS in LMXB
in quiescence. 

We stress that pairing of the bound protons and neutrons within the clusters, as well 
as the proton shell effects are crucial to get a correct 
$Q_{\rm tot}=\int_{\rm crust}{\rm d}r {\rm d}Q/{\rm d}r =
\int_{\rm crust}{\rm d}P {\rm d}Q/{\rm d}P$. Without these
effects only pycnonuclear fusion could generate heat. Then  the many-body  model is the Extended Thomas-Fermi one \cite{Fantina2018}, and $Q_{\rm tot}$ decreases  to one  third  ($\sim 0.5$~MeV) of the actual value \cite{Fantina2018}. While pairing for bound nucleons is so important, the 
effect of the superfluidity of unbound neutrons  still  remains to be considered. 
\vskip 3mm
{\bf Acknowledgements}
 This  work was supported in part by the National Science Centre, Poland, grant 2018/29/B/ST9/02013.   

\end{document}